\documentclass[journal,letterpaper]{IEEEtran_nssmic}
\usepackage{graphicx}  
\begin{document}
\title{Thick Pixelated CZT Detectors With Isolated Steering Grids}
\author{I.~Jung*$^1$\thanks{*Ira Jung: jung@physics.wustl.edu}\thanks{$^1$Department of Physics, Washington University, St. Louis, MO 63130, USA}, A.~B.~Garson$^1$, J.~S.~Perkins$^1$, H.~Krawczynski$^1$,  
J.~Matteson$^2$, R.~T.~Skelton$^2$\thanks{$^2$University of California, San Diego, 9500 Gilman Dr., La Jolla, CA 92075, USA}, A.~Burger$^3$, M.~Groza$^3$\thanks{$^3$ Department of Physics, Fisk University, Nashville, TN 37208, USA}}
\maketitle
\begin{abstract}
We explore the possibility to improve the performance of 0.5 cm thick 
Cadmium Zinc Telluride (CZT) detectors with the help of steering grids 
on the anode side of the detectors. Steering grids can improve the 
energy resolution of CZT detectors by enhancing the small pixel effect; 
furthermore, they can increase their detection efficiency by steering 
electrons to the anode pixels which otherwise would drift to the 
area between pixels. Previously, the benefit of steering grids 
had been compromised by additional noise associated with currents 
between the steering grids and the anode pixels. We use thin film 
deposition techniques to isolate the steering grid from the 
CZT substrate by a 150 nm thick layer of the isolator Al$_2$O$_3$.
While the thin layer does not affect the beneficial effect of 
the steering grid on the weighting potentials and the electric field
inside the detector, it suppresses the currents between the
steering grid and the anode pixels. In this contribution, we present 
first results from a 2$\times$2$\times$0.5~cm$^3$ CZT detector 
with 8$\times$8 pixels that we tested before and after deposition 
of an isolated steering grid. The steering grid improves the 662 keV 
energy resolution of the detector by a factor of 1.3 (from about
2\% to about 1.5\%), while not reducing the detection efficiency. 
To gain further insights into the detector response in the region 
between pixels, we measured energy spectra with a collimated $^{137}$Cs 
source. 
The collimator measurements can be used to enhance our understanding
of energy spectra measured under flood illumination of the detectors.
\end{abstract}

\begin{keywords}
CdZnTe, CZT detector, steering grid, photolithography
\end{keywords}
\vspace{-0.3cm}
\section{Introduction}
\PARstart{C}{admium} Zinc Telluride (CZT) is rapidly coming of age as a detector material for photons from a few keV to a few MeV.  It offers superior spatial and energy resolution compared to scintillators; and it is more economical and compact compared to Ge, particularly considering that CZT does not require cryogenic cooling.
\\
CZT is a compound semiconductor, with a band gap between 1.5 and 2.2 eV, depending on the Zn/Cd fraction.  This band gap allows room-temperature operation.  Its high average Z of 50 and high density contribute to effective stopping of photons.  One of its limitations is a poor hole mobility  $(\mu_h\:\tau_h=\rm(0.2-5)\cdot10^{-5}cm^2/V)$ and trapping.   Advanced electrode designs, including pixilation,
crossed strips, and steering electrodes, mitigate this by virtue of the ``small pixel effect'' \cite{Barret95,Luke95}.  
The best energy resolutions are achieved by combining small-pixel detector designs with
corrections for any residual dependence of the induced signals on the depth of the interaction
(DOI) 
of the primary photons. In practice, the DOI can be estimated by measuring the 
timing delay between cathode rise and anode rise, \cite{Kalem02,Zhang04}, or 
from the anode to cathode signal ratio (e.g.\ \cite{Kraw04}).

\begin{figure} 
\vspace{-0.4cm}
\centering
\includegraphics[width=8cm]{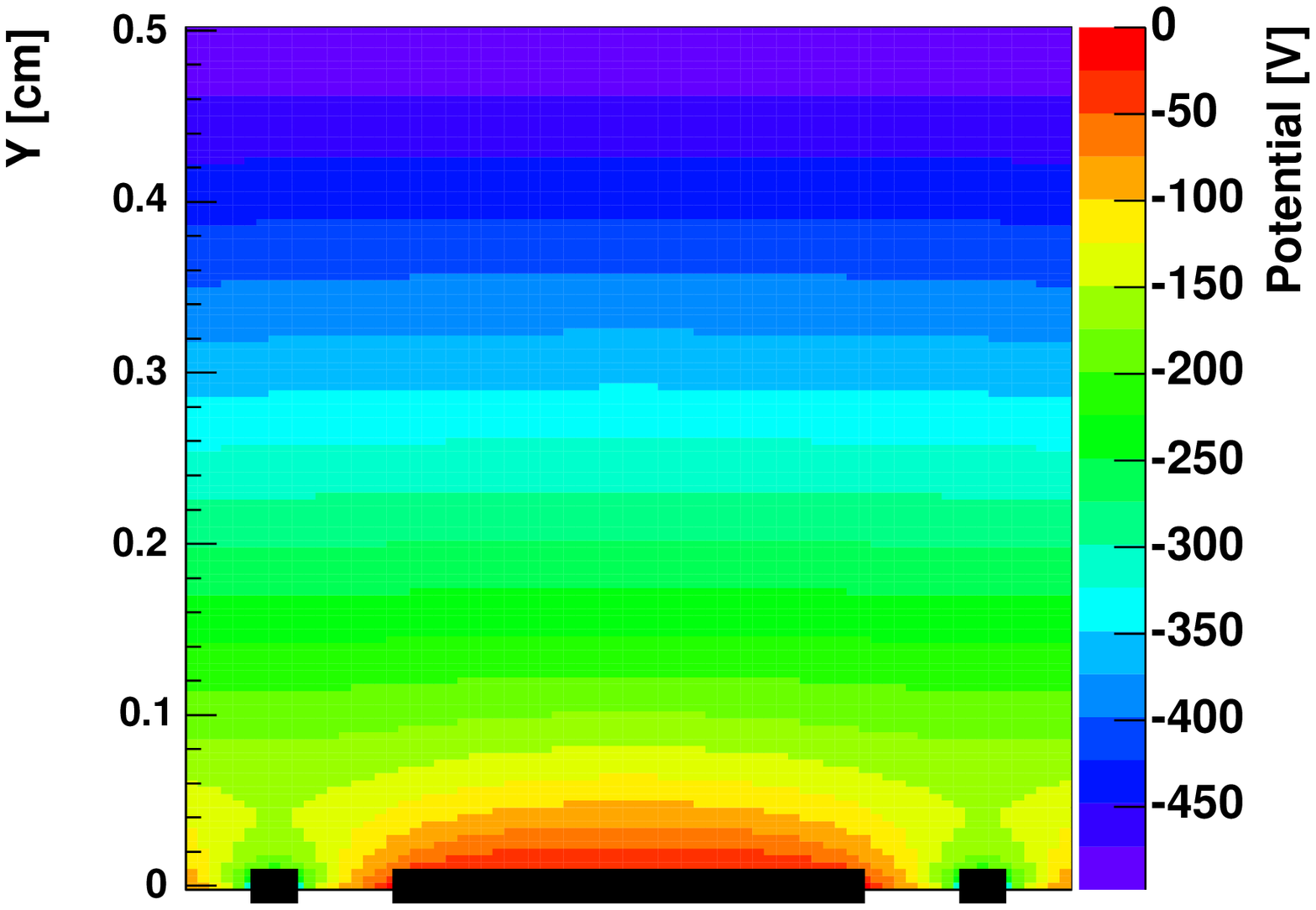}
\vspace*{-0.6cm}
\caption{Potential distribution of a central pixel in a 
0.5~cm thick detector with pixel pitch of 0.24 mm, pixel width of 
0.16 mm and a steering grid width of 0.016 mm from a 3-D detector 
simulation. While the anode pixels were held at ground, the cathode 
was biased at -500~V and the steering grid at -300~V. The 3-D Possion 
solver was developed at Washington University by S.\ Komarov.}
\label{potential}
\end{figure}

Since the use of small anode contacts leaves gaps to which charge can drift and escape collection \cite{bolotnikov}, 
steering grids biased somewhat below anode potential have been used to steer the charges to the anodes. 
Figure \ref{potential} shows the results from a 3-D detector simulation and illustrates the potential distribution 
in such a detector when the grid is biased at -300~V relative to the pixels.
 
\begin{figure} \vspace{-0.2cm}
\centering
\includegraphics[width=5.5cm]{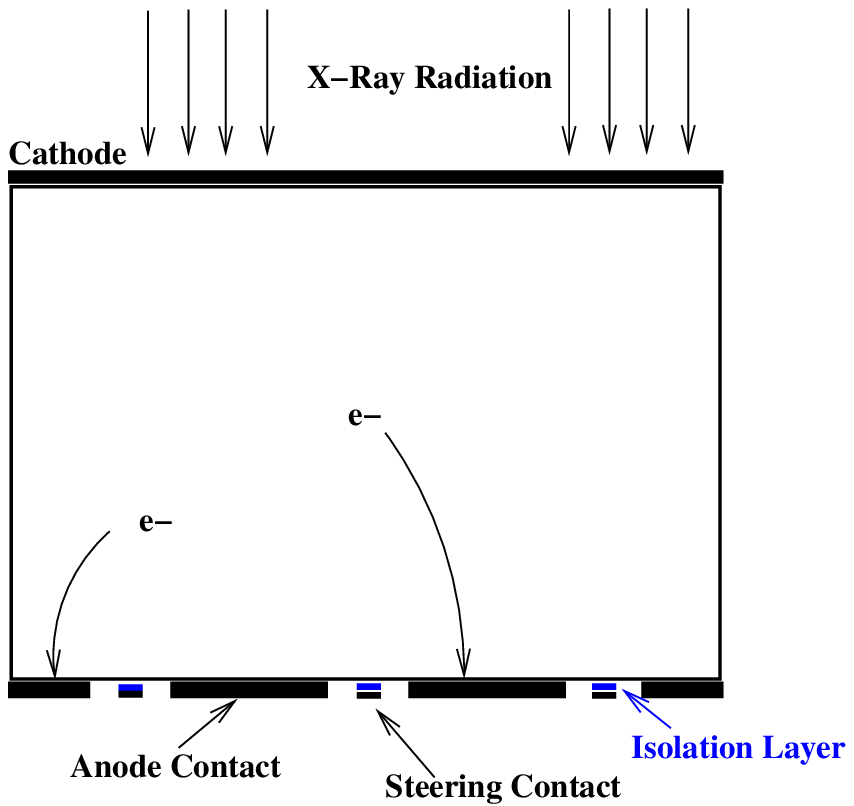}
\caption{Sketch of a CZT detector with planar cathode contact, pixelated anode contacts
and a steering grid isolated from the CZT substrate by a high-resistivity isolation layer.
We show schematically that the steering grid can steer electrons generated in the 
volume below adjacent pixels to the pixels.}
\label{Layer}
\end{figure}

CZT detectors with steering grids encounter two difficulties. First, the surface resistivity of CZT is much lower than the
bulk resistivity ($\sim$10$^9$ Ohm cm) and the grid bias voltage gives rise to currents between 
the steering grids and the pixels. The noise associated with these currents may deteriorate
the energy resolution of the detectors. 
Second, for  grid bias voltage much lower than the cathode bias voltage, a considerable 
fraction of electric field lines inside the detector connect to the steering grids. 
Thus, biased in this way steering grids  tend to collect some 
of the charge generated in the detectors and tend to reduce the detection efficiency of 
the detectors. Here we explore a novel approach that uses steering grids that are 
isolated from the CZT bulk material by a thin isolation layer (Fig.\ \ref{Layer}).
In the following we briefly outline our technique of fabricating the detectors;
subsequently we describe our measurement equipment and the results from testing the 
detectors with flood illumination and with a collimated X-ray beam.
In the following, all energy resolutions are full width half maximum (FWHM) 
resolutions. 
\section{Results}
\subsection{Detector fabrication}
The studies use modified Horizontal Bridgeman CZT from the company Orbotech 
\cite{Orbotech}. 
We fabricated several detectors. On some we used
the In pixels deposited by the Orbotech. On others we deposited the pixels 
ourselves using e-beam evaporation through a mask. 
The results shown in the following are from detector ``D1'' 
with pixels deposited 
by Orbotech and an isolated steering grid deposited in our laboratory. 

\begin{figure}
\centering
\includegraphics[width=4.5cm]{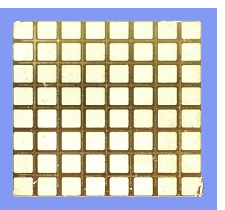}
\caption{Pixelated 2.0$\times$2.0$\times$0.5~cm$^3$ CZT detector  with steering grid. A high-resistivity Al$_2$O$_3$
film isolates the steering grid from the CZT substrate.}
\label{GridSmallBWN}
\end{figure}

The detectors have a volume of 2.0$\times$2.0$\times$0.5~cm$^3$ and are 
contacted with a planar In cathode and 8$\times$8 In pixels with a pitch of 0.25~cm and a pixel width of 0.16~cm. 
After evaluating the performance of the detectors, the isolated steering grids were
deposited using standard photolithographic techniques.
We used the photoresist S1813 and the developer Cd-30 from the company Shipley \cite{shipley}. 
We optimized the processing parameters based on a series of empirical tests 
varying the prebake and softbake temperatures and durations, 
the exposure time, and the development time.
The optimization has been carried through at Washington University in St. Louis and
at Fisk University. 
After protecting the pre-deposited pixels with photoresist, 
first a 150 nm thick isolation layer, and subsequently a 200 nm thick grid were deposited at 
$\sim$10$^{-7}$ Torr with an electron beam evaporator. We used Al$_2$O$_3$ as 
isolation material because of its very high resistivity ($>10^{14}$ Ohm cm), and excellent 
mechanical properties. Almost any metal with good sticking properties could be used 
for the steering grid. We chose Ti because of its low price and relative ease 
with which it can be deposited.
The thickness of the isolation layer has still to be optimized.
The width of the steering grid is approximately 0.02~cm.
Figure \ref{GridSmallBWN} shows one of our detectors  with a steering grid. 

%
\subsection{Experimental Setup}
Several setups have been used to test the performance of the detector. 
At Washington University the pixel performance of three of the 
central pixels are measured by using flood illumination with 
a $^{137}$Cs source.
The setup reads out four channels (the three 
central anode pixels plus the cathode, all other
anode pixels are grounded) that are AC coupled to fast Amptek 250 amplifiers 
followed by a second amplifier stage. The amplified signals are digitized by a 
500 MHz oscilloscope and transferred to a PC via Ethernet. The time resolved 
readout enables us to measure the drift time of electrons through the detector with a 
resolution of 10 ns. Each event gets a time stamp. A fit of an exponential to 
a histogram of the times between successive events after deadtime correction,
 is used to calculate
the detection rate. Care was taken to locate the source always at 
the same location above the detector, so that the measured rates can be
used to estimate the detection efficiency of the detectors.
The detector is mounted by using gold plated
pogo-pins to contact the anode, the steering grid and the cathode. 
The cathode is negatively biased, and the anode pixels are held at ground.
The steering grid was biased at -30~V, -60~V, $-120$~V and -200~V. 
The electronic noise of our test set-up has been measured before each detector 
evaluation and lies between 5~keV and 10~keV. We used a Keithley picoammeter 
to measure cathode-pixel, steering grid-pixel, and pixel-pixel IV curves.
A more detailed description of the test-equipment as well as previous
results on Orbotech detectors have been given in \cite{Kraw04,Perk03,Jung05}. 

At the University of California in San Diego the detector response has been
studied with a collimated gamma-ray source. The 7.5 cm thick 
tungsten collimator has a tapered hole. The hole diameter is 0.02 cm 
at the source and 0.05 cm at the detector end of the collimator. 
The CZT cathode is 0.10 cm away from the collimator.
The collimator position is controlled by a x-y stage with 1$\mu$m 
accuracy. Eight channels can be read out, the information obtained is the 
pulse height of the signals. The FWHM noise of anode channels 
lies between 5.75~keV to 8.1~keV for cathode biases between $-100$~V and 
$-1000$~V and grid bias between $-30$~V and $-120$~V. The FWHM noise of the cathode
channel lies between 7.2~keV and 41~keV.
More information on the experimental set-up as well as previous results
can be found in \cite{Kalem02}.
\subsection{Results}
\begin{figure} 
\centering
\includegraphics[width=8cm]{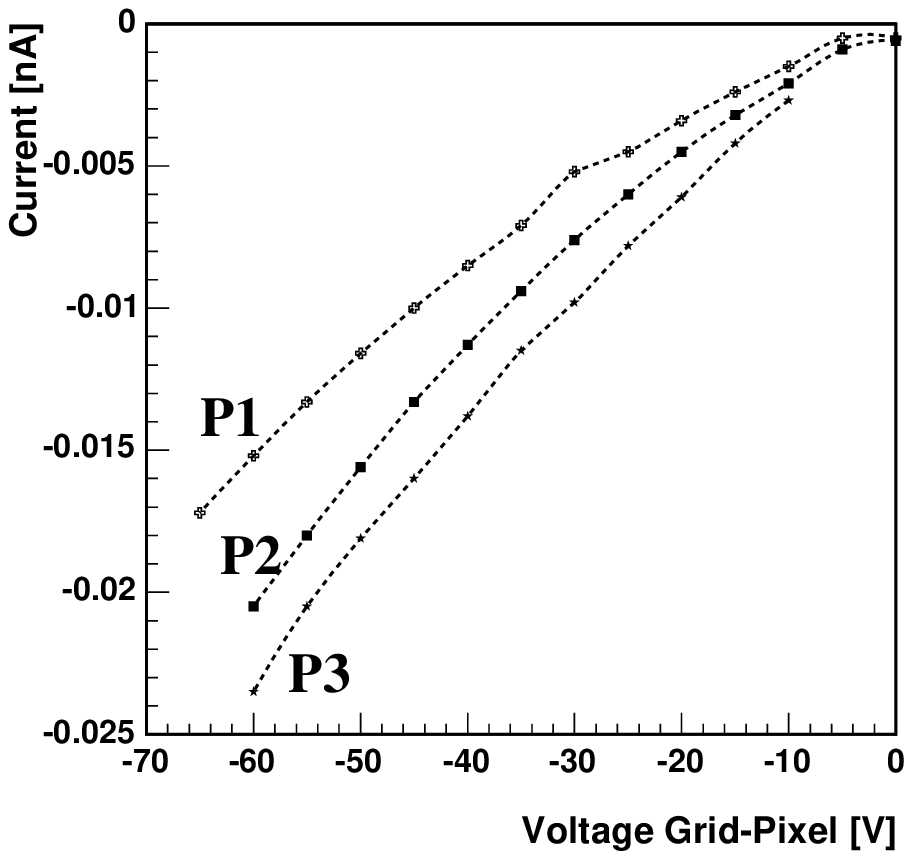}
\vspace*{-0.6cm}
\caption{The currents between the steering grid and 
three central pixels as function of relative grid bias for detector D1.}
\label{IVCurveGridPixel}
\end{figure}

In the following, we report on the results obtained with detector D1.
Figure \ref{IVCurveGridPixel} shows the current between the isolated steering grid and 
several anode pixels as function of the grid bias (pixels and cathode grounded).
For a grid-bias of -60~V grid-pixel currents per pixel below 0.25 nA were observed.
Comparing these results to pixel-pixel IV measurements of detectors without 
a steering grid, we observe that the Al$_2$O$_3$ layer results in a substantial current suppression.

\begin{figure} 
\centering
\includegraphics[width=8cm]{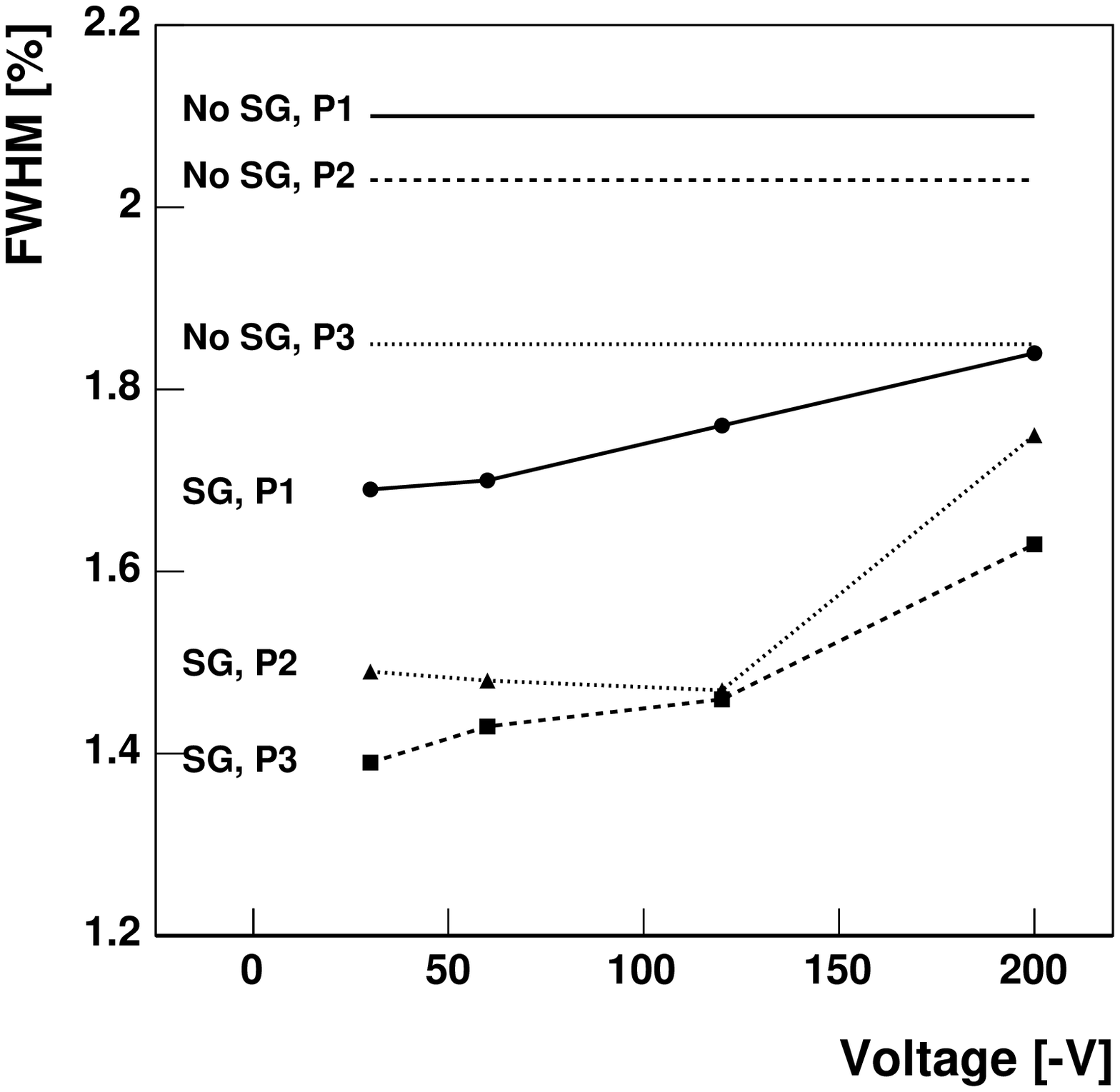}
\vspace*{-0.6cm}
\caption{Energy resolutions of the three central pixels of detector D1.
The horizontal lines show the energy resolutions before deposition of the steering grid (``No SG''), 
and the data points give the energy resolutions with the isolated steering grid at different bias
voltages (``SG'').}
\label{EnergyResolution}
\end{figure}

Figure \ref{EnergyResolution} shows the 662~keV energy resolution of the three central
pixels for flood-illuminating the detector with a $^{137}$Cs radioactive source.
Before depositing the steering grid, the three pixels gave energy resolutions 
of 2.10\%, 2.03\% and 1.85\%. Biasing the steering grid at -30~V, the performance
of the same three pixels improved to 1.69\%, 1.49\% and 1.39\%, respectively.
The energy resolutions changed little for steering grid voltages between 0~V and
-200~V. For biases $<$-200~V, the detector performance deteriorated significantly.

Before and after deposition of the steering grid, we obtained detection rates
that are identical within the statistical errors. Mantaining the same detection 
efficiency while substantially improving the energy resolution of the 
detectors is a very encouraging result.
\begin{figure}[h] 
\centering
\includegraphics[width=8cm]{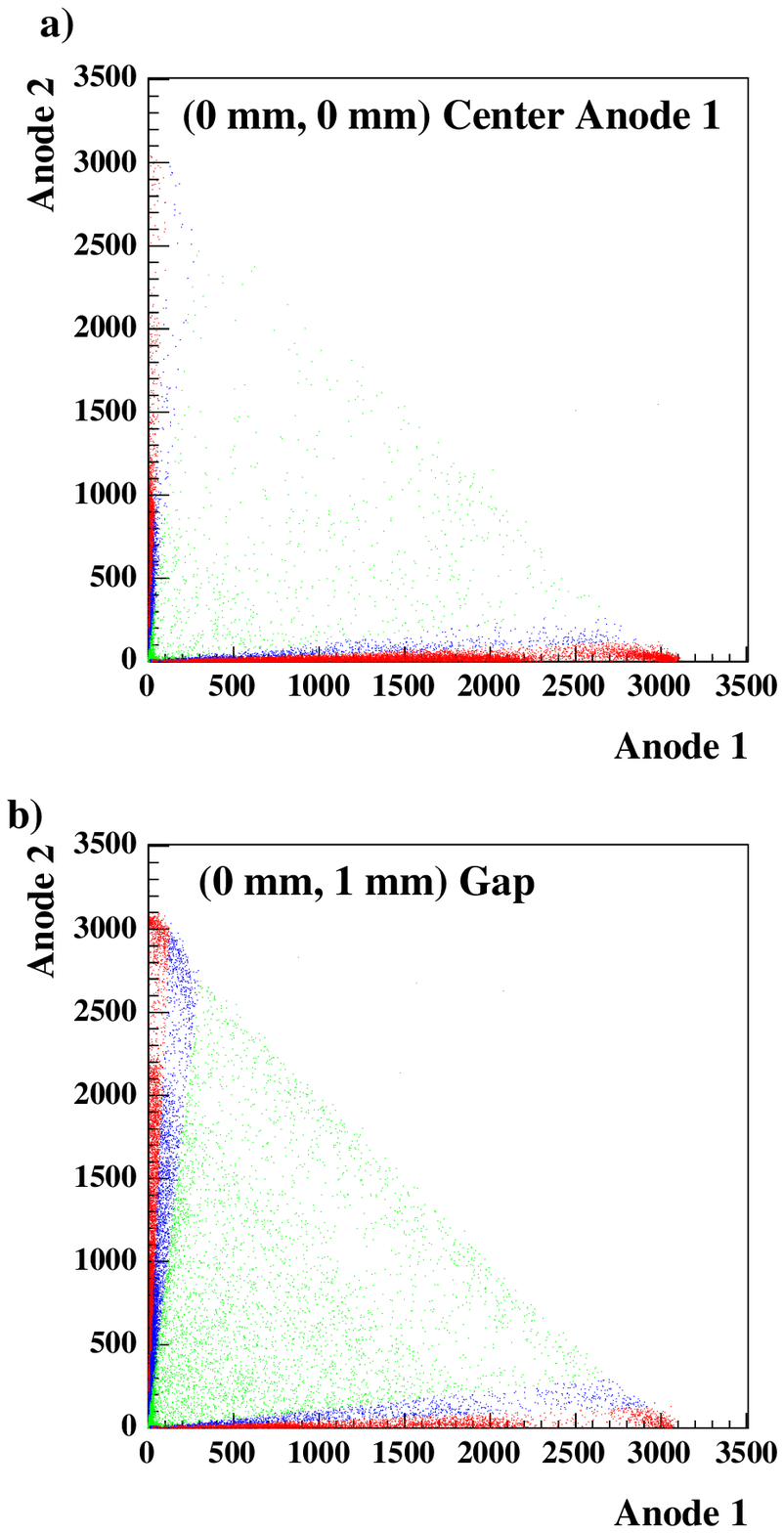}
\vspace*{-0.3cm}
\caption{Scatter plot of anode 1 versus anode 2 signals at two different 
positions of the collimated $^{137}$Cs beam: a) 
at the center of anode 1 (0 mm, 0mm) and 
b) in the middle between anode 1 and anode 2. The different colors correspond 
to different sharing ratios. 
Ref for a sharing ratio between 0-0.04 and 0.96-1.0, blue for 0.04-0.1 and 0.9-0.96
and green for 0.1-0.9.}
\label{A1vsA2}
\end{figure}

In order to understand the detector performance for energy depositions 
in the detector volume below adjacent pixels, we scanned
the detector D1 with a collimated 0.05 cm diameter $^{137}$Cs source. 
The scan with a step size of 0.125 cm started at the center of 
one pixel and ended at the center of an adjacent pixel. The detector
pixels were held at ground, the cathode was biased at -500~V, and 
the steering grid at -60~V. The electronic noise for the anode 
and cathode channels was 7~keV and 20~keV, respectively.
Here we show first results from the scans of the detector with
an isolated steering grid and their initial interpretation.
Note that we obtained qualitatively very similar results for  detectors with 
and without steering grid. More detailed studies aiming 
at studying the effect of the steering grid are in preparation.

Figure \ref{A1vsA2} shows that the pixel signals depend strongly on the 
position of the radioactive beam. Illuminating the center of ``pixel 1'' 
charge sharing is negligible and almost all events are only detected
by pixel 1 (Fig. \ref{A1vsA2}a). The small fraction (2\%) of photo-effect 
events (around channel 3040) detected with pixel 2 are most likely
662~keV photons that made it through the collimator walls and hit 
pixel 2 directly. 
In the following, we call the ratio of the charge collected with pixel 
1 divided by
the charge collected by pixels~1 and 2 the sharing ratio. 
We measured the percentage of events with a sharing ratio between 
$>$10\% and $<$90\% 
using a  cut on the sum of the two pixel signals ($>1500$ channels).
With the collimator centered at pixel 1 this fraction of 
events with substantial charge sharing was $\sim$10\%, at the position 
(0 mm, 0.5 mm) 
$\sim$25\% and in the middle between the pixels $\sim$40\% (Fig. \ref{A1vsA2}b).

\begin{figure} 
\centering
\includegraphics[width=8cm]{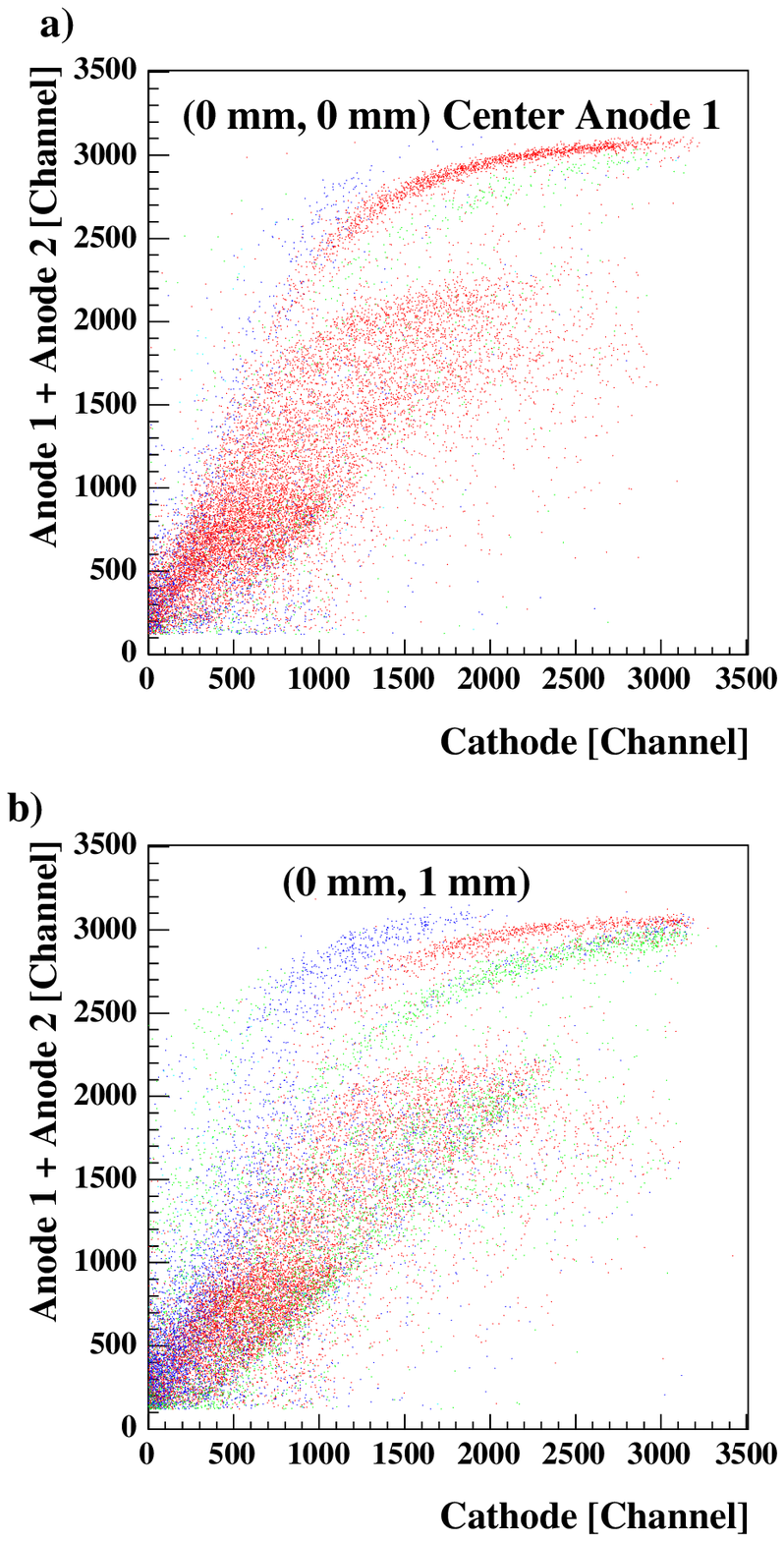}
\vspace*{-0.3cm}
\caption{Scatter plot of the sum of the signals at anodes 1 and 2 versus the
cathode signal. Only signals exceeding the threshold of 120 channels were 
included in the summation. The two plots relate to different positions
of the collimator equipped with a $^{137}$Cs source: a) 
at the center of pixel 1, b) in the middle of the gap between anodes 1 and 2.}
\label{SumAnodevsCathode}
\end{figure}

\begin{figure} 
\centering
\includegraphics[width=8cm]{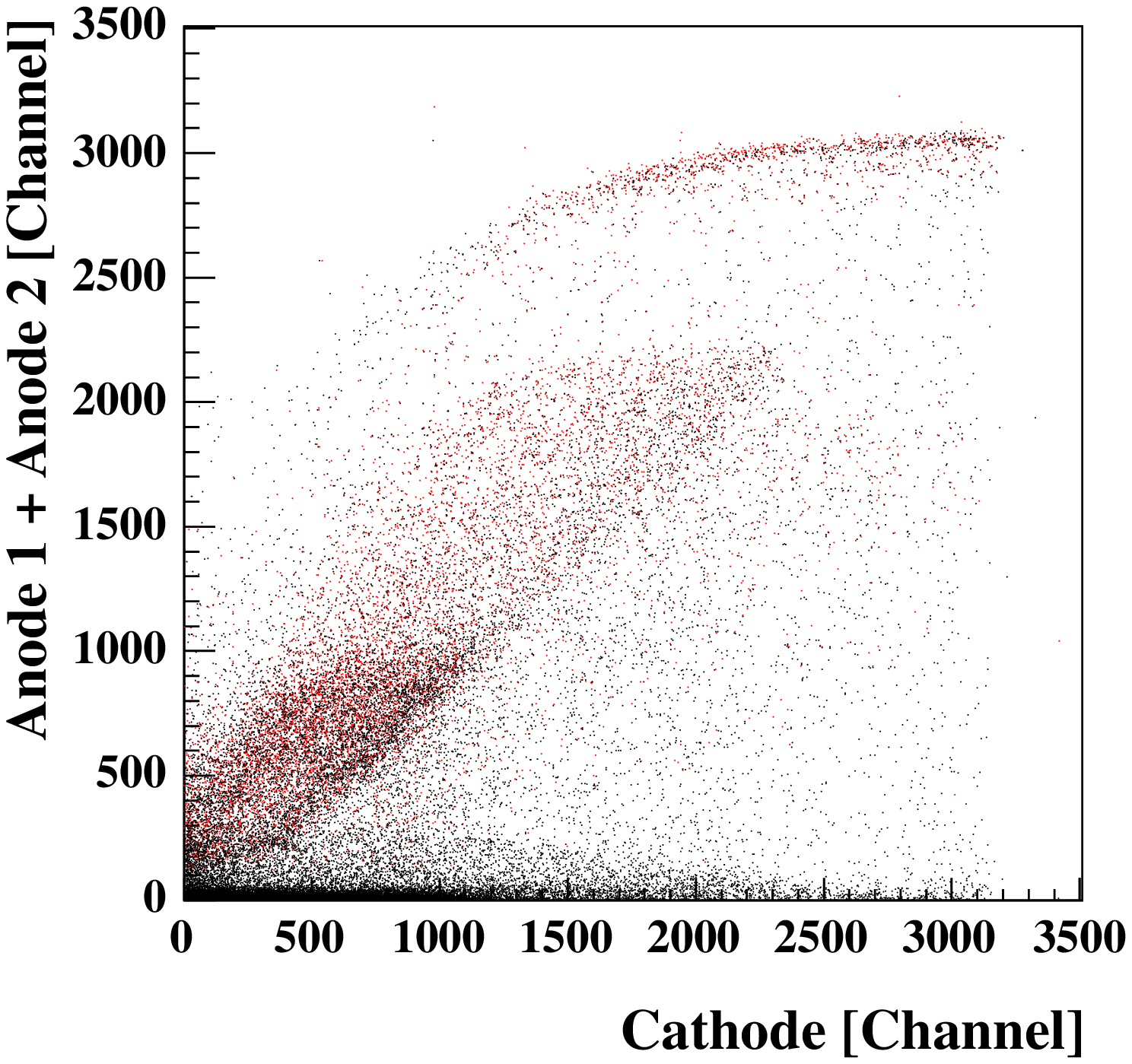}
\caption{For the collimator position at the center of 
the gap, the summed  anode signals  are shown as a function of the cathode
signal for events with sharing ratios between 0\%-4\% and 96\%-100\% (red markers). Overlayed in black, the anode~1 signal distribution
is shown for the same collimator position. }
\label{Anode1AndSum}
\end{figure}

We examined the same data in more detail. The sum of the anode signals 
of pixels~1 and 2 was plotted as a function of the cathode signal 
for the collimated beam pointing at the center of pixel 1 
(Fig.~\ref{SumAnodevsCathode}a) and at the location
between pixels 1 and 2 (Fig.~\ref{SumAnodevsCathode}b) . Only signals which exceed a threshold of 120 channels were 
included in the analysis. The different colors in Fig. \ref{SumAnodevsCathode}
show events with different sharing ratios (red: 0-4\% and 
96-100\%, blue: 4\%-10\% and 90\%-96\% and in green 10\%-90\%).
When the collimator was pointed at the region between the two pixels,
three different populations contribute to the photo-peak event line. 
The events with approximately no charge sharing (red) (in the following called line 1) 
exhibit a similar anode to cathode dependency as the events 
taken when the collimated beam pointed at the center of pixel 1  (Fig. \ref{Anode1AndSum}).
Most events with sharing ratios between 4\%-10\% and 90\%-96\% 
exhibit relatively small cathode signals and the summed anode signals 
lie above the ones with no charge sharing on a line, called ``line 2'' in the
following.
For the events with charge sharing ratios between 10\% and 90\%, 
the photo-peak line (``line 3'') is shifted to lower summed anode values.
In the following, we refer to the three photopeak lines in Fig. 
\ref{SumAnodevsCathode} as lines 1, 2 and 3.

Line 3 shows a strong dependency on the depth of interaction.
A possible explanation is lost charge in the gap/steering grid region. 
The charge loss depends 
on the depth of interaction \cite{Kalem02}. 
After energy deposition in the detector the electrons undergo diffusion. 
Neglecting the electron charge, the diffusion can be described by the Fick's equation $D\nabla^{2}M = \delta M/\delta t$. (D is the diffusivity, for CZT $\sim$26 cm$^{2}$/s) (\cite{Kalem02}). The one dimensional solution for the 
concentration M(x,t) at 
position x and time t for a delta function initial concentration M$_{0}\delta(x_{0},0)$ at position $x_{0}$
is $M(x,t)=M_{0}/\sqrt{4\pi Dt}\exp{(-(x-x_{0})^{2}/(4Dt))}$.
If one makes a rough estimate, that all events at line 3 interact 
in the middle of the gap and that a region of 18$\mu$m exactly between 
the pixels, steals charge 
from the signal, line 1 can be reproduced. 
To fully understand this effect, further investigations are needed. 
Detailed simulations and measurements are planned.  

\begin{figure} 
\centering
\includegraphics[width=8cm]{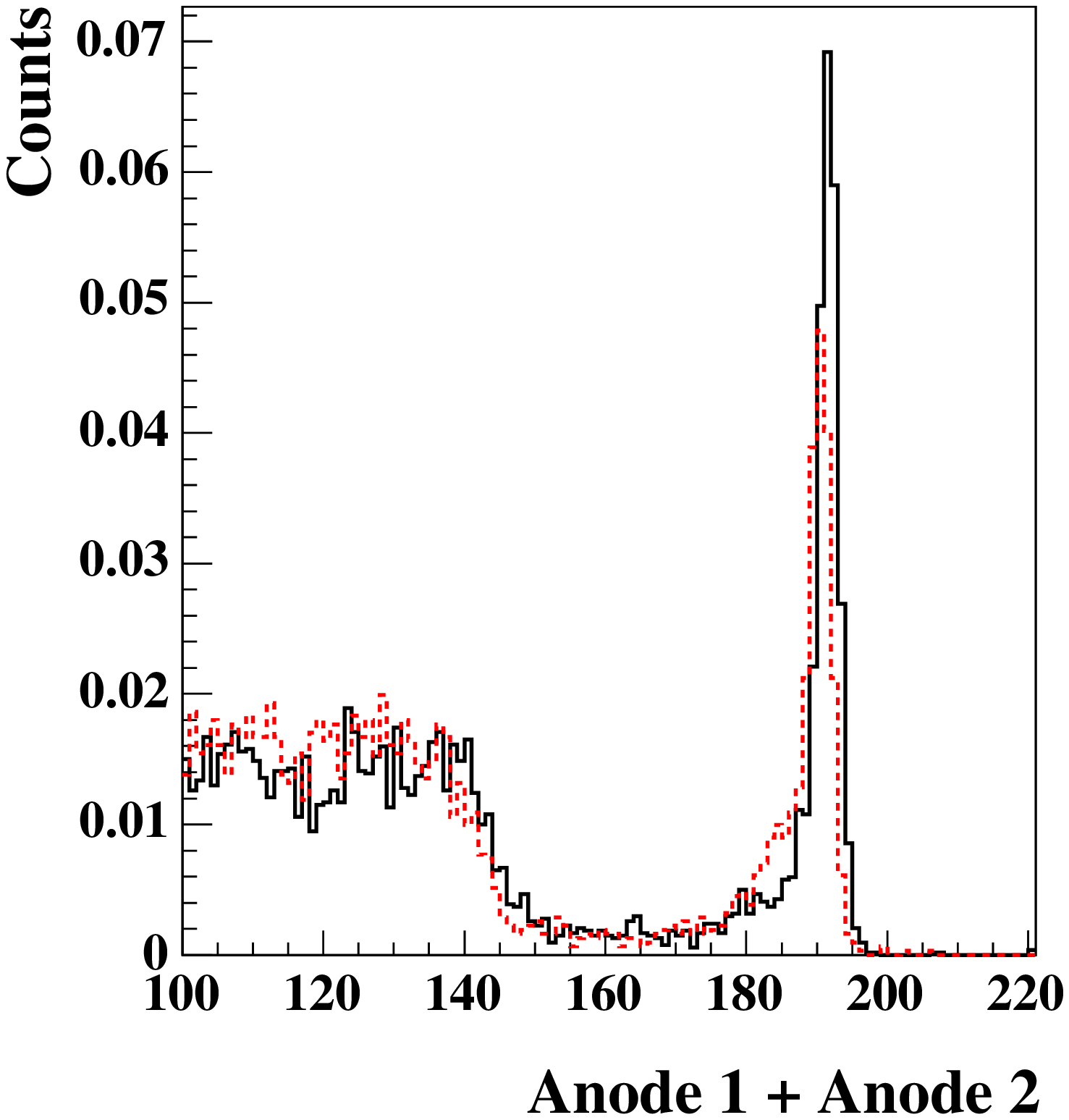}
\caption{Spectra for two different collimator positions. The black, solid line shows the spectra for the collimator centered at anode 1 (1.7\% energy resolution), the red, dashed line 
for the collimator at the pixel gap (2.1\% energy resolution). Both spectra
include corrections  for three 
different sharing ratios: a) 0\%-4\% and 96\%-100\%, b) 4\%-10\% and 90\%-96\% and c) 10\%-90\%.}
\label{correct}
\end{figure}

Events with small sharing ratios (4\%-10\% and 90\%-96\%) are mostly 
found in line 2. These are probably events interacting 
near the anode side and near the 
pixel edges. In this area, a signal can be induced in the adjacent 
pixel that increases the summed signal. 
Because the different lines are dependent on the sharing ratio,
this is an effect which  can be corrected for. 
In Fig. \ref{correct} the spectra for the collimator position at
anode 1 and for the collimator position at the gap between 
anode 1 and anode 2 is shown. Both spectra are corrected for the 
effect mentioned above. The obtained  energy resolutions are 
1.7\% at anode 1 and 2.1\% at the gap.
\\
One remainig question is the behaviour of the summed anode signal vs. cathode 
signal as a function of different grid biases. For biases 
equal to the pixel voltage and for a grid-bias of -30~V  
the number of charge sharing events is strongly suppressed compared 
to a grid-bias of -60~V.
The number of events in the region 
between 2560 channels and 3200 channels (the photopeak range) 
are 1062, 1846 and 2910 for bias 0~V, -30~V and -60~V respectively. 
The steering grid  works effectively and increases the number of 
detected photopeak events by a factor of almost 3, for events 
in the gap between pixels.

\section{Conclusion}
We fabricated pixelated CZT detectors with steering grids. As an innovation, 
we used photolithography and thin film deposition methods to fabricate detectors
with steering grids that are isolated from the CZT substrates by thin-film Al$_2$O$_3$ layers.
We presented first results obtained from testing the detectors at 662~keV.
The results are extremely encouraging: the Al$_2$O$_3$ layer seems to
reduce the grid-pixel currents and the steering grid seems to improve the
energy resolution of the detector by a factor of 1.3. 
While the grid biased at -30~V did not reduce the detection efficiency, when
using only the information from one pixel. Using the information
from adjacent pixels by summing their signals, the steering grid even 
increases the number of events reconstructed in the photopeak. 

With the aim to achieve a good understanding of the performance of the detector
with and without steering grids, we have started to scan the detector response 
with a collimated X-ray beam. The results show a very complex behavior of 
the detector for primary energy depositions in the volume below adjacent pixels. We have shown, that the sharing ratio can be used to identify different
classes of events. It should be possible to use the sharing ratio
to correct inter-pixel events, even when the detector is flood illuminated.
We still have to test this method. 
We plan to achieve further progress by  complementing the measurements
with detailed 3-D models of the detector.

Our future work will concentrate on a systematic study of the dependence 
of the detector response on the steering grid voltage, the dependence of the 
grid-pixel currents on the thickness of the Al$_2$O$_3$ isolation layer, 
and the relative performance of different isolation materials.
\subsection*{Acknowledgements}
This work has been supported by NASA under contracts NNG04WC17G and NNG04GD70G, and the NSF/HRD grant no.\ 0420516 (CREST) and by DOE National Nuclear Security Administration of Nonproliferation Research and Engineering NA-22 under grants DE-FG07-04ID14555 and DE-FG52-05NA27036.
We thank S.\ Komarov and L.\ Sobotka for the joint work on CZT detectors
and for access to their 3-D detector simulation code. We thank Orbotech, especially Y.\ Raab, A.\ Shani and U.\ El-Hanany for fruitful discussions.

\end{document}